\documentclass[conference]{IEEEtran}
\IEEEoverridecommandlockouts
\usepackage{cite}
\usepackage{amsmath,amssymb,amsfonts}
\usepackage{algorithmic}
\usepackage{graphicx}
\usepackage{textcomp}
\usepackage{xcolor}
\usepackage[hyphens]{url}
\usepackage[hidelinks]{hyperref}
\def\BibTeX{{\rm B\kern-.05em{\sc i\kern-.025em b}\kern-.08em
    T\kern-.1667em\lower.7ex\hbox{E}\kern-.125emX}}
    
\begin{document}

\title{A Conversational Agentic Interface to Physics-Based Household Digital Twins for Residential Energy Decision Support}

\author{\IEEEauthorblockN{Costas Mylonas}
\IEEEauthorblockA{\textit{Energy Digitalisation Group} \\
\textit{UBITECH}\\
Athens, Greece \\
kmylonas@ubitech.eu}
\and
\IEEEauthorblockN{Titos Georgoulakis}
\IEEEauthorblockA{\textit{Energy Digitalisation Group} \\
\textit{UBITECH}\\
Athens, Greece \\
tgeorgoulakis@ubitech.eu}
\and
\IEEEauthorblockN{Magda Foti}
\IEEEauthorblockA{\textit{Energy Digitalisation Group} \\
\textit{UBITECH}\\
Athens, Greece \\
mfoti@ubitech.eu}
}

\maketitle

\begin{abstract}
Multiple actors around residential energy systems require accessible decision-support tools: homeowners and tenants for dwelling-level retrofit choices, consultants and municipal planners for building and district-level intervention assessment, and retailers and aggregators for estimating residential flexibility and coordinating distributed energy resources. However, existing pathways remain limited, since professional audits are costly and static, rule-of-thumb estimates lack household specificity, and high-fidelity simulation tools require specialized expertise. This paper presents a conversational agentic framework that makes physics-based household energy simulation accessible through natural language interaction. The proposed system integrates a Household Digital Twin (HDT), built on GridLAB-D and exposed through a REST-based microservices architecture, with a two-tier large language model (LLM) agentic layer that translates user requests into structured, schema-compliant simulation payloads. To improve reliability, the architecture combines intent routing, a domain-specific knowledge base, deterministic post-processing of simulation outputs, and tool-governed execution policies. The system is evaluated on a curated dataset of 45 prompts with increasing complexity, covering multiple households, seasons, and override scenarios. Results show 100\% schema conformance, 96.1\% field-level F1, 90.4\% value accuracy, and a 95.6\% end-to-end simulation success rate. The findings indicate that conversational agentic interfaces can substantially lower the usability barrier of physics-based household digital twins while preserving the reliability required for residential energy decision support.
\end{abstract}


% \maketitle

\section{Introduction} \label{sec:introduction}

Residential buildings are central to the energy transition because they are both major energy consumers and emerging sources of demand-side flexibility. In the European Union, households accounted for 26.2\% of final energy consumption in 2023 \cite{eurostat2026households}. As heat pumps, rooftop photovoltaics, batteries, electric vehicles, and dynamic tariffs spread, actors across residential energy systems need tools that quantify the energy, comfort, cost, and flexibility implications of decisions before implementation: homeowners and tenants for retrofit and electrification choices, consultants and municipal planners for building- and district-scale interventions, and retailers and aggregators for assessing distributed flexibility for tariff design, portfolio management, and demand response.

In practice, however, accessible and reliable decision support remains limited. Homeowner-oriented retrofit tools are useful for initial screening, but many rely on empirical methods, pre-simulated databases, or simplified normative calculations, and often provide limited personalization and weak consideration of user preferences \cite{seddiki2021retrofit}. At the other end of the spectrum, physics-based simulation tools can represent building envelope characteristics, HVAC systems, appliances, and weather with much higher fidelity, but they require substantial expertise in model formulation, parameterization, and output interpretation. This creates a persistent gap between the analytical power of advanced simulation and the accessibility required by the actors who need to use such analyses in practice.

Building digital twins offer a promising foundation for narrowing this gap. Recent reviews show strong momentum in building digital twins for operational energy efficiency, monitoring, and decision support \cite{cespedescubides2024bdt}. At the same time, research on human-building interaction has begun to argue for the democratization of digital twins, that is, transforming them from specialist back-end models into interfaces through which occupants and other users can explore building behaviour more directly \cite{lee2023democratization}. Nevertheless, most existing building digital twin applications still emphasize on visualization, monitoring or expert-driven analysis, rather than conversational authoring of counterfactual residential energy scenarios that can support retrofit assessment, electrification planning, or flexibility-oriented decision making.

In parallel, natural-language interfaces, Large Language Models (LLMs), and agentic AI are becoming increasingly relevant to energy and building applications. The authors of \cite{hulsmann2021nli} showed that natural-language interaction can make complex energy system models more accessible to non-expert users, while \cite{jiang2024eplusllm} proposed EPlus-LLM to translate natural-language descriptions into EnergyPlus models. More directly relevant to our framework, ReAct combines step-by-step reasoning with external tool use \cite{yao2023react}, Toolformer studies how models decide when and how to invoke external APIs \cite{schick2023toolformer}, and AutoGen demonstrates multi-agent task decomposition through conversable agents \cite{wu2024autogen}. Related building-energy work is also beginning to adopt such ideas, including a platform-agnostic schema for reusable LLM agents in building energy analysis \cite{zhang2025agentschema} and a multi-agent workflow for automated building energy modeling and calibration \cite{lu2025data2bem}. However, these advances also highlight a key challenge: natural-language convenience alone is not sufficient for trustworthy decision support. When user requests must be converted into executable simulation payloads, the interface must preserve schema compliance, physical plausibility, and numerical reliability.

This paper addresses that challenge through a conversational agentic interface for a physics-based Household Digital Twin (HDT) built on GridLAB-D. GridLAB-D is an open-source under Linux foundation for energy, agent-based simulation framework for smart grids \cite{chassin2014gridlabd}, and its residential end-use module represents household thermal behaviour through an equivalent thermal parameter (ETP) formulation \cite{taylor2008residential}. Building on this simulation core, we develop a two-tier conversational architecture that translates natural-language requests into structured, simulation-ready scenarios for residential energy analysis. The contribution of the paper is threefold: first, we present the design of an HDT framework that couples a GridLAB-D-based residential digital twin with a conversational interface; second, we introduce an agentic architecture that combines intent routing, domain-grounded payload generation, and deterministic post-processing to improve reliability; and third, we evaluate the approach using a structured prompt set of increasing complexity to assess structural correctness, semantic accuracy, robustness, and end-to-end simulation success.

The remainder of the paper is organized as follows. Section \ref{sec:proposed_framework} presents the proposed framework, including the household digital twin engine and the conversational agentic layer. Section \ref{sec:evaluation_methodology} describes the evaluation methodology. Section \ref{sec:results} reports and discusses the experimental results. Section \ref{sec:conclusion} concludes the paper and outlines future work.

\section{Proposed Framework}
\label{sec:proposed_framework}

\begin{figure}[t]
    \centering
    \includegraphics[width=\columnwidth]{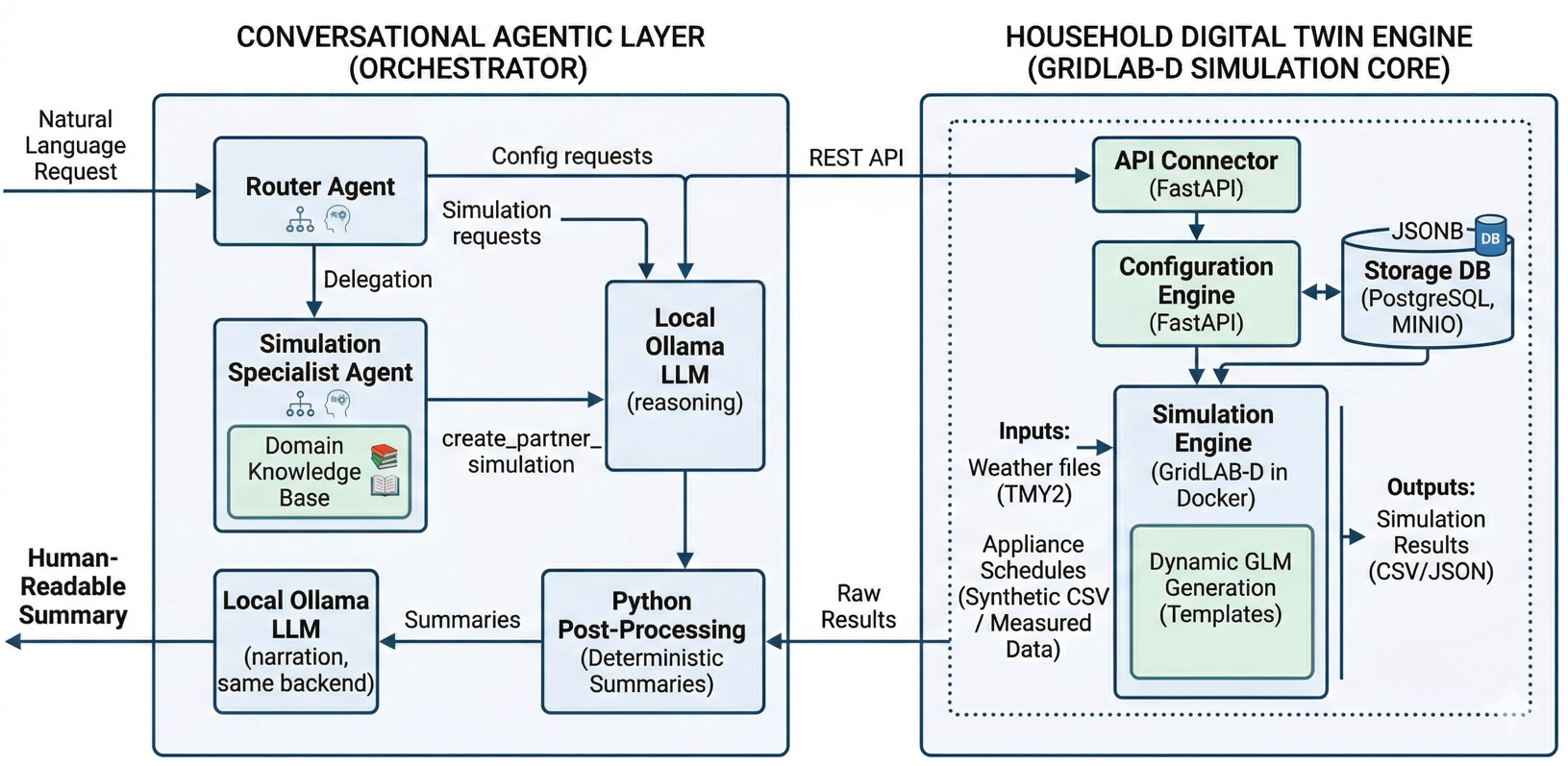}
    \caption{Architecture of the proposed HDT framework.}
    \label{fig:hdt_architecture}
\end{figure}

The proposed Household Digital Twin (HDT) framework, shown in Fig.~\ref{fig:hdt_architecture}, combines a conversational agentic layer with a GridLAB-D-based simulation engine connected through a REST API. The simulation backend provides executable household energy scenarios, while the agentic layer translates natural-language requests into structured API calls and returns grounded summaries of the results. This separation allows the system to preserve simulation fidelity while making it accessible to non-expert users.

\subsection{Household Digital Twin Engine}

The HDT engine is implemented as a containerized microservices application built around GridLAB-D. It models the household components most relevant to residential energy analysis, including the building envelope, HVAC system, domestic hot water, individual appliances, and optional assets such as photovoltaic generation, electric-vehicle charging, and battery storage.

The backend is organized into three services. The \emph{Configuration Engine} manages household definitions and stores the static parameters of each dwelling, such as geometry, insulation levels, orientation, thermostat settings, and appliance metadata. The \emph{Simulation Engine} generates and executes GridLAB-D scenarios based on these household configurations. The \emph{API Connector} exposes the functionality through REST endpoints, enabling external applications and the conversational interface to create, run, and retrieve simulations and their respective results.

For each request, the selected household configuration is combined with the simulation horizon, sampling interval, requested output properties, and any user-defined overrides. These inputs are translated into a GridLAB-D model file through template-based generation. Simulation outputs are exported as time series and stored together with scenario metadata for traceability and reproducibility.

To support households without complete appliance-level measurements, the HDT also includes a synthetic data generation pipeline. Representative appliance cycles are adapted to the target device characteristics and converted into schedules that can be injected into the simulation. When measured appliance data are available, they take precedence over the synthetic schedules. Weather conditions are provided through site-specific typical meteorological year files, ensuring that all scenarios remain grounded in local climatic conditions.

\subsection{Conversational Agentic Layer}

The conversational layer is exposed through a chat-oriented REST interface. Users express requests in natural language, and the agent interprets them, assembles valid simulation payloads, invokes the corresponding backend endpoints, and returns human-readable summaries. The agent therefore acts as an orchestrator between the user and the HDT engine rather than as a simulation component itself.

The system adopts a two-tier design. A \emph{Router Agent} receives each user request and determines whether it concerns household configuration management or simulation. Configuration-related actions can be handled directly, whereas simulation-related requests are delegated to a specialized \emph{Simulation Specialist Agent}. This separation confines schema-sensitive payload construction to the component that is explicitly designed for that task.

The Simulation Specialist Agent constructs complete simulation payloads using a domain knowledge base loaded at startup. This knowledge base captures the API structure, required fields, admissible parameter ranges, default behaviors, and supported household components. As a result, the specialist operates under explicit schema guidance rather than generating payloads in an unconstrained manner.

To improve reliability, numerical reporting is separated from language generation. After a simulation is executed, deterministic post-processing routines compute the relevant summary values from the raw outputs before the final response is produced. The language model receives only these verified summaries and converts them into a natural-language report of the simulation outcomes. If that step fails, the system falls back to a deterministic template-based response. In addition, write operations require explicit approval, and repeated tool calls within the same request are cached to avoid unnecessary backend execution.

\subsection{Reliability Mechanisms}

A central design objective of the proposed framework is to constrain the agentic AI layer so that it remains useful for decision support without behaving as an unconstrained conversational model. This is achieved through several complementary mechanisms. First, task separation limits the responsibilities of each agent: the Router Agent handles intent routing, while the Simulation Specialist Agent is solely responsible for schema-sensitive payload construction. Second, the specialist operates with an explicit domain knowledge base that defines the admissible API structure, supported fields, and parameter constraints. Third, tool execution is governed by policies that require explicit approval for write operations and avoid redundant executions through in-request caching. Finally, the generation of user-facing summaries is separated from numerical computation, so that all reported quantitative values are derived deterministically from simulation outputs before being expressed in natural language. Together, these mechanisms reduce hallucination risk, improve reproducibility, and make the conversational interface more suitable for trustworthy residential energy analysis.

\subsection{Illustrative Interaction Example}

An example request is: "Simulate the HVAC load for the ATHENS.2 household in July with the cooling setpoint lowered to 24$^\circ$C and return hourly results.". The Router Agent first identifies the request as simulation-related and delegates it to the Simulation Specialist Agent. The specialist then identifies the requested household and maps the time horizon, sampling interval, requested outputs, and setpoint override into a structured simulation payload. After execution by the HDT backend, deterministic post-processing extracts the relevant summary statistics, which are finally returned to the user in natural language. This workflow illustrates how the framework converts conversational requests into executable and interpretable simulation scenarios without requiring the user to interact directly with the underlying API schema.

\section{Evaluation Methodology} \label{sec:evaluation_methodology}

The evaluation examines whether the proposed conversational agent can reliably translate natural-language household energy requests into valid, schema-compliant, and executable simulation payloads, and how this reliability changes with request complexity. All experiments were performed with the same local LLM setup used by the framework, namely Mistral Nemo 12B served through Ollama, configured with temperature 0 for deterministic responses and a 16{,}384-token context window to accommodate the prompt, tool metadata, conversation history, and intermediate results within a single inference call.

\subsection{Evaluation Dataset}

A curated dataset of 45 test cases was constructed for the evaluation. Each case consists of a natural-language prompt and a ground-truth JSON payload representing the correct simulation request. The dataset is organized into three tiers of increasing complexity, with 15 cases per tier.

Tier 1 (\emph{Baseline}) tests the basic mapping from natural language to simulation request, including household identification, simulation horizon, sampling interval, and requested output properties, without parameter modifications. Tier 2 (\emph{Single Override}) introduces one modification category per case, such as an HVAC setpoint change, a building-envelope adjustment, a nested object parameter, or an appliance schedule. Tier 3 (\emph{Multi-Category}) combines multiple modifications within a single prompt, requiring the agent to assemble payloads spanning several sections of the schema. The most complex cases include multiple override fields, nested parameters, appliance schedules, and a large set of requested outputs.

The dataset covers seven household configurations across Greece, Ireland, Spain and Croatia, and includes summer, winter, and spring/autumn scenarios. Simulation horizons range from single days to full months, while sampling intervals include 5-minute, 15-minute, and hourly resolutions. This design ensures that the evaluation reflects a realistic range of residential simulation requests rather than a narrow set of simplified examples.

\subsection{Evaluation Metrics}

Performance is assessed across five dimensions. \emph{Structural correctness} measures whether the generated output is well formed and includes all mandatory top-level fields of the simulation payload, such as the household reference, simulation horizon, sampling interval, and output specification; this is reported through schema conformance and action format correctness. \emph{Payload accuracy} measures agreement with the ground truth using field-level precision, recall, and F1 over flattened dot-separated field paths, together with value accuracy and exact match rate.

\emph{Semantic accuracy} captures task-specific correctness and includes household identification, override inclusion accuracy, appliance inclusion accuracy, output property precision and recall, temporal accuracy, and interval accuracy. \emph{Robustness} measures failure modes through hallucination rate, defined as the presence of invalid or unknown fields, and constraint compliance, defined as adherence to domain rules such as physically valid thermostat settings. Finally, \emph{operational performance} is measured through simulation success rate, i.e., whether the generated payload is accepted and executed by the HDT backend, and mean end-to-end latency.

This protocol distinguishes between payloads that are merely well formed, payloads that are semantically correct, and payloads that are fully executable in the simulation backend, providing a more informative assessment than exact-match accuracy alone.

\section{Results and Discussion} \label{sec:results}

Table~\ref{tab:overall_results} reports the overall performance across all 45 evaluation cases. The proposed framework achieves 100\% schema conformance and 100\% action format correctness, showing that the agent consistently produces well-formed requests and invokes the correct backend action. At the payload level, it reaches 96.1\% field F1 and 92.\% value accuracy, while end-to-end simulation success reaches 95.6\%. These results indicate that the framework is structurally reliable and operationally robust. The exact-match rate is lower, at 37.8\%, because this metric counts a case as correct only when the generated payload matches the ground-truth payload in every field, including time settings, output selections, and override parameters.

\begin{table}[t]
\caption{Overall evaluation results ($n=45$).}
\label{tab:overall_results}
\centering
\scriptsize
\begin{tabular}{p{1.35cm}p{2.75cm}r}
\hline
Dimension & Metric & Value \\
\hline
Structural & Schema conformance & 100\% \\
& Action format correctness & 100\% \\
Payload & Field F1 & 96.1\% \\
& Value accuracy & 92.6\% \\
& Exact match rate & 37.9\% \\
Semantic & Override inclusion accuracy & 100\% \\
& Appliance inclusion accuracy & 95.6\% \\
& Output property precision & 100\% \\
& Output property recall & 77.3\% \\
& Temporal accuracy & 100\% \\
& Interval accuracy & 100\% \\
Robustness & Constraint compliance & 100\% \\
& Hallucination rate & 4.4\% \\
Operational & Simulation success rate & 95.6\% \\
& Mean latency & 9,785 ms \\
\hline
\end{tabular}
\end{table}

The semantic metrics show that the main limitations lie in omission rather than invalid generation. Override inclusion accuracy is 100\%, appliance inclusion accuracy is 95.6\%, and output property precision is 100\%, meaning that the agent rarely introduces unsupported content. Temporal accuracy and interval accuracy both reach 100\%, demonstrating that the agent reliably resolves simulation time horizons and sampling rates from natural language descriptions. However, output property recall is 77.3\%, indicating that errors are mainly associated with missing expected outputs rather than incorrect generation.

Table~\ref{tab:per_tier_results} shows that performance degrades gradually with prompt complexity. Field F1 decreases from 1.000 in Tier 1 to 0.969 in Tier 2 and 0.913 in Tier 3, while simulation success remains high across all tiers. Value accuracy is stable across tier at approximately 92-93\%. Hallucination is 0\% in the first two tiers and rises to 13.3\% only in Tier 3, where prompts combine multiple override categories. Mean latency follows the same trend, increasing from 8.8--9.5~s in the simpler tiers to 11.1~s in Tier 3.

\begin{table}[t]
\caption{Per-tier evaluation results ($n=15$ per tier).}
\label{tab:per_tier_results}
\centering
\scriptsize
\begin{tabular}{lccc}
\hline
Metric & T1 & T2 & T3 \\
\hline
Field F1 & 1.000 & 0.969 & 0.913 \\
Value accuracy & 92.5\% & 92.7\% & 92.7\% \\
Output prop. recall & 79.0\% & 71.7\% & 81.3\% \\
Temporal accuracy & 100\% & 100\% & 100\% \\
Hallucination rate & 0\% & 0\% & 13.3\% \\
Simulation success & 93.3\% & 100\% & 93.3\% \\
Mean latency (ms) & 9,456 & 8,784 & 11,116 \\
\hline
\end{tabular}
\end{table}

A closer inspection of the errors reveals two recurring failure modes. First, some cases omit output variables that are implicit in the ground truth, such as total household load, which lowers recall without introducing invalid properties. Second, the hallucination cases in Tier 3 stem from appliance schedules being placed in the wrong payload section, indicating a structural confusion rather than arbitrary generation. Notably, the agent never selects a nonexistent output property and never violates physical constraints such as thermostat ordering, confirming that the knowledge-base grounding is effective at preventing invalid content.

The results show that the proposed framework is most reliable in the parts of the workflow that are explicitly constrained by the architecture. In particular, the combination of two-tier task separation, schema-guided payload construction, and deterministic post-processing appears to be responsible for the 100\% schema conformance and the high simulation success rate. This suggests that the main benefit of the agentic design is not open-ended reasoning, but controlled orchestration of a physics-based backend under clear structural constraints.

At the same time, the remaining errors reveal where the current design is still limited. The omission of implicitly expected outputs indicates that correct payload construction is not only a matter of schema validity, but also of modelling conventions and user expectations. These findings suggest that future improvements should focus on better task decomposition, stronger intermediate validation, and enriched knowledge-base conventions for implicit output selection, rather than solely on larger language models.

From an application perspective, the observed performance is already sufficient to support exploratory residential energy analysis in which users compare scenarios, test override configurations, and inspect likely impacts on demand and comfort. However, for deployment in decision-critical workflows, the agentic layer should evolve toward more explicit verification, clarification, and uncertainty handling. In this sense, the evaluation highlights both the promise and the current boundary of conversational agentic interfaces for household digital twins: they are highly effective at making simulation workflows accessible, but they still require additional safeguards when user intent is incomplete or implicitly specified.

\section{Conclusion}
\label{sec:conclusion}

This paper presented a conversational agentic framework that makes a physics-based Household Digital Twin (HDT) accessible through natural-language interaction. The proposed approach combines a GridLAB-D-based simulation engine with a two-tier agentic layer that routes user requests, constructs schema-compliant simulation payloads, and returns grounded summaries through deterministic post-processing. By coupling a physics-based digital twin with a conversational interface, the framework reduces the technical barrier of household energy simulation while preserving the fidelity required for residential energy analysis.

The evaluation demonstrated that the framework is structurally reliable and operationally effective across a diverse set of residential simulation requests. The agent achieved 100\% schema conformance, 100\% temporal and interval accuracy and 95.6\% end-to-end simulation success. Remaining errors are confined to the omission of implicitly expected output properties and, in the most complex multi-category prompts, occasional structural confusion in appliance schedule placement, neither of which involves unsupported or physically invalid payload generation. These results indicate that conversational agentic interfaces can substantially lower the usability barrier of household digital twins while maintaining the reliability needed for retrofit assessment, electrification analysis, and flexibility-oriented decision support.

Future work will focus on extending the agentic AI layer beyond the current router--specialist design. Promising directions include adding explicit planning and verification agents, incorporating dynamic retrieval and tool selection mechanisms, enabling clarification-driven interaction for ambiguous requests, and introducing memory-aware multi-turn reasoning for iterative scenario exploration. Another important direction is the development of self-checking and uncertainty-aware agent behaviors that can detect low-confidence payloads before execution and trigger validation or user confirmation steps. These extensions can further improve robustness, adaptability, and trustworthiness in conversational energy simulation workflows.

\section*{Acknowledgment}

This work was supported by European Union’s funded Project DIGITISE [grant number 101160671].

\bibliographystyle{IEEEtran}
\bibliography{references}

\end{document}